\documentclass[aps,pra,twocolumn,superscriptaddress]{revtex4-2}

\usepackage{amsmath,amssymb}
\usepackage{graphicx}
\usepackage{bm}
\usepackage{physics} 

\graphicspath{{./figures/}}

\allowdisplaybreaks[4]

\newcommand{\RIKEN}{RIKEN Center for Quantum Computing, 2-1 Hirosawa, Wako-shi, Saitama 351-0198, Japan}
\newcommand{\UTokyo}{Department of Applied Physics, School of Engineering, The University of Tokyo, 7-3-1 Hongo, Bunkyo-ku, Tokyo 113-8656, Japan}

\begin{document}

\title{Configuration design of multimode Gaussian operations on continuous-variable quad-rail lattice cluster states}

\author{Jun-ichi Yoshikawa}
\affiliation{\RIKEN}
\author{Warit Asavanant}
\affiliation{\UTokyo}
\author{Hironari Nagayoshi}
\affiliation{\UTokyo}
\author{Atsushi Sakaguchi}
\affiliation{\RIKEN}
\author{Shota Yokoyama}
\affiliation{\RIKEN}
\author{Hidehiro Yonezawa}
\affiliation{\RIKEN}
\author{Akira Furusawa}
\affiliation{\RIKEN}
\affiliation{\UTokyo}

\email{jun-ichi.yoshikawa@riken.jp}
\date{\today}

\begin{abstract}
Continuous-variable quad-rail lattice cluster states enable flexible quantum circuit design on their two-dimensional structure. 
However, how to combine basic operations on the quad-rail lattice cluster state to realize multimode operations has not been deeply discussed. 
Here we show a concrete configuration design to efficiently implement beamsplitter network operations on the cluster state. 
Furthermore, combining the beamsplitter networks, a configuration design of multimode Gaussian unitary operations is also shown. 
It is theoretically known that the Gaussian operations are sufficient for universal quantum computation if appropriate non-Gaussian states are injected. 
Our results are fundamentally important for utilizing the flexible quad-rail lattice cluster states for computations. 
\end{abstract}

\maketitle

\section{Introduction}
\label{sec:introduction}

Measurement-based quantum computation (MBQC) is a promising approach instead of unitary gate-based implementation~\cite{Raussendorf.prl2001,Briegel.natphys2009,Menicucci.prl2006}. 
MBQC utilizes cluster states as universal resource states, which are multipartite entangled states with the quantum correlation structure expressed by a graph. 
Large-scale cluster states, which lead to large-scale quantum computation, can be created in continuous-variable (CV) regime with optical setup by employing time-domain multiplexing or frequency-domain multiplexing techniques~\cite{Menicucci.pra2011,Menicucci.prl2008,Alexander.pra2016}. 
Recently there are many experimental demonstrations of such large-scale cluster states~\cite{Yokoyama.natphoton2013,Chen.prl2014,Roslund.natphoton2014,Yoshikawa.aplphoton2016,Asavanant.science2019,Larsen.science2019,Asavanant.prapplied2021}.

Various graph structures are considered for the CV cluster states. 
Among them, two-dimensional quad-rail lattice (QRL) cluster states are advantageous because flexible wiring of quantum information flow is possible~\cite{Alexander.pra2016}. 
The concept of macronode and micronode is introduced to the QRL cluster graph, and a calculation step of MBQC is shown to be a pair of generalized quantum teleportations from one macronode to neighboring macronodes. 
The generalized quantum teleportations apply various unitary operations on the teleported quantum states, depending on the measurement bases. 
The destination of the teleportation is also freely chosen from the neighboring macronodes by the measurement bases, which is the reason for the flexibility in designing quantum wires on the QRL cluster state. 
It is theoretically known that a combination of basic unitary operations on the QRL cluster state available with homodyne measurements (such as beamsplitter, phase shift, and squeezing operations) enables an arbitrary Gaussian unitary operation. 
However, concrete configuration design on the QRL cluster state to make general multimode Gaussian operations has not been deeply discussed. 
In the previous work, the measurement-based linear optics on the QRL cluster states was proposed~\cite{Alexander.prl2017}; however, the considered circuit design was not so efficient because redundant phase-shift components increased the number of teleportations.

Here we show a concrete design to implement a general beamsplitter network on the QRL cluster state efficiently. 
The configuration is based on the famous decomposition of a beamsplitter network by Reck \textit{et al.}~\cite{Reck.prl1994}, and we show a triangle-shaped configuration enables a general beamsplitter network operation without introducing redundancies. 
We explicitly show how the decomposition by Reck \textit{et al.} is converted to the decomposition that fits to the QRL cluster state. 
In addition, we also provide another configuration of a beamsplitter network on the QRL cluster state, based on the decomposition by Clements \textit{et al.}~\cite{Clements.optica2016}. 
Roughly speaking, these configurations decrease the number of teleportations by half, compared to the previous design in Ref.~\cite{Alexander.prl2017}: While a macronode for a phase shift operation is inserted between every pair of macronodes for beamsplitter operations in Ref.~\cite{Alexander.prl2017}, our design do not need that. 
Furthermore, considering the Bloch-Messiah reduction~\cite{Braunstein.pra2005}, we also show a configuration to realize an arbitrary Gaussian unitary operation, which consists of the two triangle-shaped beamsplitter networks sandwiching single-mode squeezing macronodes. 
Although more efficient configurations may exist for general Gaussian operations, our Bloch-Messiah reduction-based configuration guarantees the maximum required size of configuration on the QRL cluster state. 
On the other hand, we may find a subgroup of general Gaussian unitaries that can be implemented more efficiently than the Bloch-Messiah reduction-based configuration. 
In fact, we find a nontrivial example of such a subgroup, which we call multimode shear operations. 
We show that the multimode shear operations can be implemented with a single triangular configuration (the same size as the Reck-based beamsplitter network configuration), which is much more efficient than the Bloch-Messiah reduction-based configuration (containing two Reck-based beamsplitter network configurations). 
These results are fundamentally important when the flexible QRL cluster states are utilized for computations. 
It is worth noting that the Gaussian operations are sufficient for universal quantum computation if appropriate non-Gaussian states are injected~\cite{Gottesman.pra2001,Baragiola.prl2019,Konno.prresearch2021}. 
Externally created non-Gaussian states can be injected by optical switches or instead by quantum teleportations~\cite{Asavanant.pra2023}.

The paper is structured as follows. 
In Sec.~\ref{sec:continusou_variables}, the basics and notations regarding CVs are summarized. 
In Sec.~\ref{sec:QRL_cluster}, the QRL cluster state is described, and the concepts of macronodes and micronodes, local and distributed, are explained. 
Note that our definition of the QRL cluster state is different from the original one~\cite{Alexander.pra2016} up to local phase redefinitions. 
(Such redefinition is often adopted in experiments~\cite{Yokoyama.natphoton2013,Chen.prl2014,Roslund.natphoton2014,Yoshikawa.aplphoton2016,Asavanant.science2019,Larsen.science2019,Asavanant.prapplied2021}.) 
In Sec.~\ref{sec:macronode}, the single-step Gaussian operation consuming a macronode of the QRL cluster state is described, which is interpreted as generalized quantum teleportations. 
In Sec.~\ref{sec:beamsplitter_network}, the implementation of general beamsplitter networks on the QRL cluster state is described. 
In Sec.~\ref{sec:Gaussian}, extension to general Gaussian unitaries is described. 
Sec.~\ref{sec:beamsplitter_network} and Sec.~\ref{sec:Gaussian} are the main results of this paper. 
Sec.~\ref{sec:conclusion} is the conclusion.

\section{Basics of continuous variables}
\label{sec:continusou_variables}

Continuous-variable systems are described as a tensor product of harmonic oscillators. 
The position and momentum of the $j$-th harmonic oscillator, denoted by the operators $\hat{x}_j$ and $\hat{p}_j$, respectively, are conjugate variables, satisfying the canonical commutation relation $[\hat{x}_j,\hat{p}_k]=i\hbar\delta_{jk}$ where $i$ is the imaginary unit and $\delta_{jk}$ is the Kronecker delta. 
We take $\hbar =1$ in the following. 
Each harmonic oscillator is often called a qumode.

For quantum optics, the generalized position $\hat{x}_j$ and the generalized momentum $\hat{p}_j$ correspond to quadratures. 
Annihilation operators $\hat{a}_j=(1/\sqrt{2})(\hat{x}_j+i\hat{p}_j)$ and creation operators $\hat{a}_j^\dagger=(1/\sqrt{2})(\hat{x}_j-i\hat{p}_j)$, satisfying $[\hat{a}_j,\hat{a}_k^\dagger]=\delta_{jk}$, correspond to complex amplitudes. 
Here the superscript $\dagger$ denotes the Hermitian conjugate, which we use in the following for both quantum operators (expressed with hats) and matrices (without hats). 
A quadrature at an arbitrary phase, 
\begin{align}
\hat{p}_j(\theta):=\hat{p}_j\cos\theta+\hat{x}_j\sin\theta=\sqrt{2}\Im(\hat{a}_je^{i\theta}), 
\label{eq:quadrature}
\end{align}
is an observable and can be measured by employing a homodyne measurement.

\subsection{Beamsplitter network}

The action of a general beamsplitter network $\hat{U}$ on $N$ qumodes is expressed uniquely, in the Heisenberg representation, by using an $N\times N$ unitary matrix $U$ as 
\begin{align}
\hat{U}^\dagger
\begin{pmatrix}
\hat{a}_1 \\ \vdots \\ \hat{a}_N
\end{pmatrix}
\hat{U}
=
\begin{pmatrix}
\hat{U}^\dagger\hat{a}_1\hat{U} \\ \vdots \\ \hat{U}^\dagger\hat{a}_N\hat{U}
\end{pmatrix}
=
U
\begin{pmatrix}
\hat{a}_1 \\ \vdots \\ \hat{a}_N
\end{pmatrix}. 
\label{eq:beamsplitter_network}
\end{align}
We can easily check the beamsplitter network transformation conserves the total photon number $\sum_{j=1}^N \hat{n}_j = \sum_{j=1}^N \hat{a}_j^\dagger\hat{a}_j$ by using $U^\dagger U=I$, where $I$ is an identity matrix. 
A beamsplitter network is a passive optical component, and the conservation of the total photon number corresponds to the conservation of energy. 
A beamsplitter network is in general constructed by combining beamsplitters, 
\begin{align}
\hat{B}_{jk}^\dagger(\tau)\begin{pmatrix} \hat{a}_j \\ \hat{a}_k \end{pmatrix}\hat{B}_{jk}(\tau)
=
\begin{pmatrix} \cos\tau & -\sin\tau \\ \sin\tau & \cos\tau \end{pmatrix}
\begin{pmatrix} \hat{a}_j \\ \hat{a}_k \end{pmatrix}, 
\label{eq:beamsplitter}
\end{align}
and phase shifts, 
\begin{subequations}
\begin{align}
\hat{R}_j^\dagger(\theta) \hat{a}_j \hat{R}_j(\theta) = & e^{i\theta}\hat{a}_j, \\
\hat{R}_j^\dagger(\theta) \begin{pmatrix} \hat{x}_j \\ \hat{p}_j \end{pmatrix}\hat{R}_j(\theta)
= &
\begin{pmatrix} \cos\theta & -\sin\theta \\ \sin\theta & \cos\theta \end{pmatrix}
\begin{pmatrix} \hat{x}_j \\ \hat{p}_j \end{pmatrix}. 
\end{align}
\end{subequations}
Taking the real or imaginary part of Eq.~\eqref{eq:beamsplitter}, beamsplitter transformation of $(\hat{x}_j,\hat{x}_k)^T$ or $(\hat{p}_j,\hat{p}_k)^T$ is also expressed with the same matrix. 
(The superscript $T$ denotes the transpose of a matrix.) 
Half-reflecting beamsplitters frequently appear in this paper, for which we omit the argument $\tau$,  
\begin{align}
\hat{B}_{jk}^\dagger\begin{pmatrix}\hat{a}_j \\ \hat{a}_k \end{pmatrix}\hat{B}_{jk}
=\frac{1}{\sqrt{2}}\begin{pmatrix}1 & -1 \\ 1 & 1\end{pmatrix}\begin{pmatrix}\hat{a}_j \\ \hat{a}_k\end{pmatrix}. 
\label{eq:half_beamsplitter}
\end{align}
Note that $\hat{B}_{jk}^\dagger=\hat{B}_{kj}$. 
Phase-shift operations are also referred to as rotation operations.

\subsection{Gaussian unitary}

A general Gaussian unitary operation $\hat{G}$ acting on $N$ qumodes is expressed in the Heisenberg representation as a Bogoliubov transformation in the form of 
\begin{align}
\hat{G}^\dagger
\begin{pmatrix}
\hat{a}_1 \\ \vdots \\ \hat{a}_N
\end{pmatrix}
\hat{G}
=A
\begin{pmatrix}
\hat{a}_1 \\ \vdots \\ \hat{a}_N
\end{pmatrix}
+B
\begin{pmatrix}
\hat{a}_1^\dagger \\ \vdots \\ \hat{a}_N^\dagger
\end{pmatrix}
+
\begin{pmatrix}
\alpha_1 \\ \vdots \\ \alpha_N
\end{pmatrix}. 
\end{align}
Here the term $(\alpha_1 \dots \alpha_N)^T$ represents a displacement in the phase space, which is easily implemented and thus neglected in the following.

In Ref.~\cite{Braunstein.pra2005}, the Bloch-Messiah reduction of this general Gaussian unitary operation is shown as the decomposition 
\begin{align}
\hat{G} = \hat{U}\hat{S}\hat{V}^\dagger, 
\label{eq:Bloch_Messiah}
\end{align}
which is based on the singular value decomposition of the matrices, 
\begin{align}
A= & UD_AV^\dagger, & B= & UD_BV^T. 
\end{align}
Here, $\hat{U}$ and $\hat{V}$ are beamsplitter networks, $U$ and $V$ are corresponding unitary matrices, $\hat{S}=\bigotimes_j\hat{S}_j$ is single-mode squeezing operations on individual qumodes, and $D_A$ and $D_B$ are diagonal matrices satisfying, 
\begin{align}
D_A^2=D_B^2+I, \label{eq:BMR_diagonal}
\end{align}
which corresponds to $\hat{S}$. 
A single-mode squeezing operation is, 
\begin{subequations}
\begin{align}
\hat{S}_j^\dagger(r) \begin{pmatrix} \hat{x}_j \\ \hat{p}_j \end{pmatrix}\hat{S}_j(r)
= & 
\begin{pmatrix} e^r & 0 \\ 0 & e^{-r} \end{pmatrix}
\begin{pmatrix} \hat{x}_j \\ \hat{p}_j \end{pmatrix}, \\
\hat{S}_j^\dagger(r)\hat{a}_j\hat{S}_j(r) 
= & 
\hat{a}_j\cosh r + \hat{a}_j^\dagger\sinh r. \label{eq:squeezing_a}
\end{align}
\end{subequations}
Squeezing of the $\hat{x}_j$ quadrature corresponds to $r<0$, and squeezing of the $\hat{p}_j$ quadrature corresponds to $r>0$. 
Note that the relation $\cosh^2 r = \sinh^2 r +1$ of Eq.~\eqref{eq:squeezing_a} corresponds to each diagonal element of Eq.~\eqref{eq:BMR_diagonal}.

\subsection{Continuous basis}

In the state-vector space (Hilbert space) of the $j$-the qumode, the eigenstate of a quadrature operator $\hat{x}_j$ with a real eigenvalue $s\in\mathbb{R}$ is expressed as $\ket{x=s}_j$. 
This eigenstate is unphysical, however, ignoring the normalization problem, considered as the limit of the infinitely squeezed vacuum state with respect to the $\hat{x}_j$ quadrature, followed by a displacement operation by $s$ in the direction of the $\hat{x}_j$ quadrature. 
Adopting the Dirac's delta innerproduct $\braket{x=s}{x=s^\prime}=\delta(s-s^\prime)$, an arbitrary 1-mode pure state is expressed as a superposition state $\int_{\mathbb{R}}ds\,\psi(s)\ket{x=s}_j$, where $\psi(s)$ is the wave function.

A two-mode squeezed vacuum state, often called as an Einstein-Podolsky-Rosen state~\cite{Einstein.pr1935}, is a basic entangled state in CV systems, and frequently appears in this paper. 
In the limit of infinite squeezing, the two-mode squeezed vacuum state is expressed as, 
\begin{align} 
\ket{\text{TS}}_{jk}
&= \iint_{\mathbb{R}^2}dsds^\prime\,\delta(s-s^\prime)\ket{x=s}_j\ket{x=s^\prime}_k \notag\\
&= \int_{\mathbb{R}}ds\,\ket{x=s}_j\ket{x=s}_k. 
\label{eq:two-mode_squeeze}
\end{align}
This state is the simultaneous eigenstate of $\hat{x}_j-\hat{x}_k$ and $\hat{p}_j+\hat{p}_k$ with zero eigenvalues.

\section{Quad-rail lattice cluster states}
\label{sec:QRL_cluster}

\begin{figure}[t]
\centering
\includegraphics[scale=0.8]{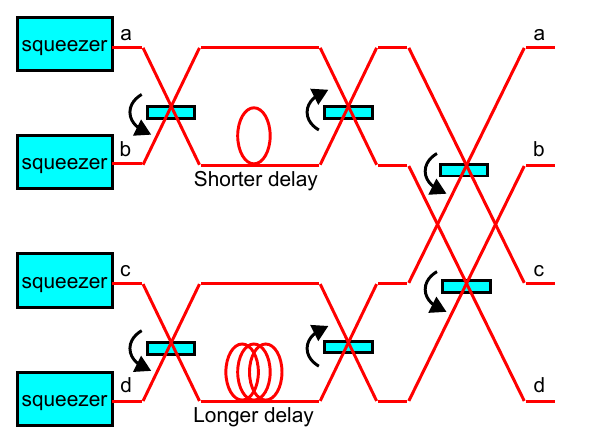}
\caption{Optical setup to generate quad-rail lattice cluster states multiplexed in the time domain. The phase factor for each beamsplitter is specified by an arrow.}
\label{fig:setup}
\end{figure}

The QRL cluster state multiplexed in the time domain is generated with an optical setup depicted in Fig.~\ref{fig:setup}. 
Squeezed light pulses are sequentially generated from four squeezers (such as optical parametric oscillators), labelled by $\ell\in\{a,b,c,d\}$, with the same intervals. 
The light pulses are specified by $(\ell,t)$, a set of the spatial label $\ell\in\{a,b,c,d\}$, corresponding to the source squeezer, and the time index $t\in\mathbb{Z}$, where $\mathbb{Z}$ denotes the set of all integers. 
(Here we are not interested in the start and end points of the cluster state in the time domain.) 
These light pulses are initially in separable squeezed states, and converted to the QRL cluster state by the interferometer in Fig.~\ref{fig:setup}. 
There are two optical delay lines in the interferometer: the upper one is shorter, which shifts the time index by one ($t\to t+1$), and the lower one is $N$ times longer than the upper, which shifts the time index by $N$ ($t\to t+N$). Here, $N$ is some large integer. 
As mentioned in the last paragraph of Sec.~\ref{sec:introduction}, the QRL cluster state we treat is different from the original theory~\cite{Alexander.pra2016} up to local phase shifts (or phase redefinitions). 
Therefore, here we fully describe the QRL cluster state. 
Of course, the capabilities of quantum computation using the QRL cluster states do not change by the local phase shifts.

\subsection{Nullifiers}

Since the structure of Gaussian entanglement is often specified by showing their nullifiers, here we describe the QRL cluster state by following step by step the transformations of nullifiers in the ideal limit of infinite squeezings.

In general, a nullifier $\hat{\mathcal{N}}$ associated with a quantum state $\ket{\psi}$ means an operator that nullifies the quantum state, $\hat{\mathcal{N}}\ket{\psi}=0$. 
Instead of expressing $\ket{\psi}$ itself in the large Hilbert space on some basis, which often becomes messy, the state $\ket{\psi}$ is also specified by a set of nullifiers. 
For Gaussian states, nullifiers can be chosen as a linear combination of quadrature operators $\hat{x}_j$ and $\hat{p}_j$ and the identity operator, and we only treat such nullifiers in the following. 
For example, a vacuum state $\ket{n=0}_j$, which satisfies $\hat{a}_j\ket{n=0}_j=0$, is expressed by the nullifier of an annihilation operator $\hat{a}_j\propto\hat{x}_j+i\hat{p}_j$. 
In particular, nullifiers for infinitely squeezed states can be chosen as a linear combination of the quadrature operators and the identity operator with real coefficients, which is an observable, e.g., $(\hat{x}_j-s)\ket{x=s}_j=0$. 
An example of entangled states is a two-mode squeezed vacuum state, for which we consider the infinitely squeezed state $\ket{\text{TS}}_{jk}$ in Eq.~\eqref{eq:two-mode_squeeze}, and the corresponding nullifiers are $\hat{x}_j-\hat{x}_k$ and $\hat{p}_j+\hat{p}_k$. 
In general, for an $N$-mode state, $N$ independent nullifiers are required to uniquely specify a single state. 
Note that since a linear combination of nullifiers is also a nullifier [$(\hat{\mathcal{N}}_1+\hat{\mathcal{N}}_2)\ket{\psi}=0$ if $\hat{\mathcal{N}}_1\ket{\psi}=0$ and $\hat{\mathcal{N}}_2\ket{\psi}=0$], nullifiers make a vector space. 
A minimum set of nullifiers to uniquely specify a quantum state is a basis of the vector space, and an arbitrary basis of the vector space can be chosen to specify the same quantum state.

When a quantum state is unitarily transformed as $\ket{\psi}\to\hat{U}\ket{\psi}$, the corresponding nullifier is transformed as $\hat{N}\to\hat{U}\hat{N}\hat{U}^\dagger$ because $(\hat{U}\hat{N}\hat{U}^\dagger)(\hat{U}\ket{\psi})=\hat{U}(\hat{N}\ket{\psi})=0$. 
Based on this, the nullifiers are converted by the optical setup in Fig.~\ref{fig:setup} as follows. 
Initially, there are four squeezed light paths. 
We suppose $\hat{p}$-squeezed states for the squeezers $a$, $c$, and $\hat{x}$-squeezed states for the squeezers $b$, $d$. 
The nullifiers to specify the sequential squeezed vacuum states are, in the limit of infinite squeezings, 
\begin{align}
\{\hat{p}_{a,t}, \hat{x}_{b,t}, \hat{p}_{c,t}, \hat{x}_{d,t}\}_{t\in\mathbb{Z}}. 
\end{align} 
First, the upper and lower pairs are combined by beamsplitters $\prod_{t\in\mathbb{Z}}\hat{B}_{cd,t}\hat{B}_{ab,t}$, resulting in pairs of two-mode squeezed states, expressed by the nullifiers, 
\begin{align}
\{ & \hat{p}_{a,t}+\hat{p}_{b,t}, \hat{x}_{b,t}-\hat{x}_{a,t}, \notag\\
& \hat{p}_{c,t}+\hat{p}_{d,t}, \hat{x}_{d,t}-\hat{x}_{c,t}\}_{t\in\mathbb{Z}}. 
\end{align}
Note that here the inverse matrix of Eq.~\eqref{eq:half_beamsplitter} is used. 
The delay lines shift the time indices as, 
\begin{align}
\{ & \hat{p}_{a,t}+\hat{p}_{b,t+1}, \hat{x}_{b,t+1}-\hat{x}_{a,t}, \notag\\ 
& \hat{p}_{c,t}+\hat{p}_{d,t+N}, \hat{x}_{d,t+N}-\hat{x}_{c,t}\}_{t\in\mathbb{Z}}. 
\label{eq:nullifier_TS_delayed}
\end{align}
Finally, they are transformed by the four beamsplitters, called a foursplitter. 
Here, the transformation of annihilation operators by the foursplitter $\hat{\mathcal{B}}_{abcd,t} := \hat{B}_{bd,t}\hat{B}_{ac,t}\hat{B}_{cd,t}^\dagger\hat{B}_{ab,t}^\dagger$ is 
\begin{align}
\hat{\mathcal{B}}_{abcd,t}^\dagger\begin{pmatrix}\hat{a}_{a,t}\\ \hat{a}_{b,t}\\ \hat{a}_{c,t}\\ \hat{a}_{d,t}\end{pmatrix}\hat{\mathcal{B}}_{abcd,t}
=\mathcal{B}\begin{pmatrix}\hat{a}_{a,t}\\ \hat{a}_{b,t}\\ \hat{a}_{c,t}\\ \hat{a}_{d,t}\end{pmatrix}, 
\label{eq:foursplitter}
\end{align}
where
\begin{align}
\mathcal{B} = & \frac{1}{2}
\begin{pmatrix}
1 & 0 & -1 & 0 \\
0 & 1 & 0 & -1 \\
1 & 0 & 1 & 0 \\
0 & 1 & 0 & 1
\end{pmatrix}
\begin{pmatrix}
1 & 1 & 0 & 0 \\
-1 & 1 & 0 & 0 \\
0 & 0 & 1 & 1 \\
0 & 0 & -1 & 1
\end{pmatrix} \notag\\
= & \frac{1}{2}
\begin{pmatrix}
1 & 1 & -1 & -1 \\
-1 & 1 & 1 & -1 \\
1 & 1 & 1 & 1 \\
-1 & 1 & -1 & 1
\end{pmatrix}. 
\end{align}
Since this foursplitter transformation is applied to all time indices $t\in\mathbb{Z}$, the overall transformation by the foursplitter is $\hat{\mathcal{B}}_{abcd}=\prod_{t\in\mathbb{Z}}\hat{\mathcal{B}}_{abcd,t}$. 
Using $\mathcal{B}^{-1}=\mathcal{B}^{T}$, the nullifiers of the resulting QRL cluster state is calculated as, 
\begin{align}
\{ & (\hat{p}_{a,t}-\hat{p}_{b,t}+\hat{p}_{c,t}-\hat{p}_{d,t}) \notag\\
& +(\hat{p}_{a,t+1}+\hat{p}_{b,t+1}+\hat{p}_{c,t+1}+\hat{p}_{d,t+1}), \notag\\
& -(\hat{x}_{a,t}-\hat{x}_{b,t}+\hat{x}_{c,t}-\hat{x}_{d,t}) \notag\\
& +(\hat{x}_{a,t+1}+\hat{x}_{b,t+1}+\hat{x}_{c,t+1}+\hat{x}_{d,t+1}), \notag\\
& (-\hat{p}_{a,t}+\hat{p}_{b,t}+\hat{p}_{c,t}-\hat{p}_{d,t}) \notag\\
& +(-\hat{p}_{a,t+N}-\hat{p}_{b,t+N}+\hat{p}_{c,t+N}+\hat{p}_{d,t+N}), \notag\\
& -(-\hat{x}_{a,t}+\hat{x}_{b,t}+\hat{x}_{c,t}-\hat{x}_{d,t}) \notag\\
& +(-\hat{x}_{a,t+N}-\hat{x}_{b,t+N}+\hat{x}_{c,t+N}+\hat{x}_{d,t+N}) \}_{t\in\mathbb{Z}}. 
\label{eq:nullifer_QRL_local}
\end{align}

As already mentioned, the QRL cluster state described here is, compared with the one in the original proposal~\cite{Alexander.pra2016}, different up to local phase shifts or equivalently, local redefinitions of quadrature operators. 
Thanks to the local phase shifts, we have obtained nullifiers where $\hat{x}$ and $\hat{p}$ quadratures are not mixed. 
Such nullifiers are often preferred by experimentalists, and the local phase shifts have been adopted in their experimental demonstrations~\cite{Yokoyama.natphoton2013,Chen.prl2014,Roslund.natphoton2014,Yoshikawa.aplphoton2016,Asavanant.science2019,Larsen.science2019,Asavanant.prapplied2021}. 
Beyond this, here we adopt the locally phase-shifted notation because the multimode shear operation in Sec.~\ref{sec:Gaussian} becomes clearer.

\subsection{Macronodes, local and distributed micronodes}

\begin{figure}[t]
\centering
\includegraphics[scale=0.8]{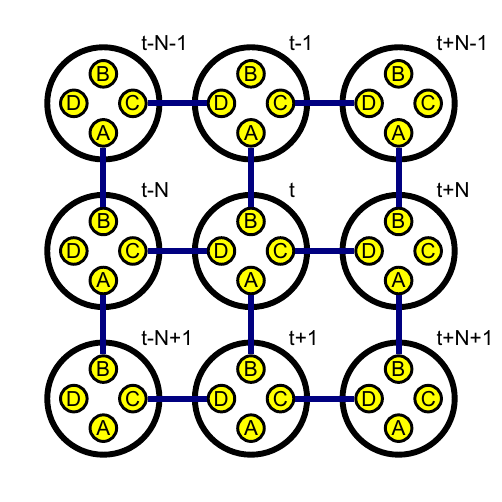}
\caption{Two-dimensional structure of the QRL cluster state based on decomposition of each macronode by distributed modes $A$, $B$, $C$, $D$ in Eq.~\eqref{eq:distributed_mode}. Edges between two distributed modes represent entanglement by two-mode squeezing.}
\label{fig:QRL_diagram}
\end{figure}

In the QRL cluster state, each quantum light pulses specified by $(\ell,t)$ with $\ell\in\{a,b,c,d\}$ and $t\in\mathbb{Z}$ is called a local micronode. 
Gathering micronodes with the same time index, we can consider a macronode $\{(a,t),(b,t),(c,t),(d,t)\}$ for each time index $t\in\mathbb{Z}$. 
The QRL cluster state has a two-dimensional structure of entanglement with respect to these macronodes, where a macronode $t$ is connected with $t-1$, $t+1$, $t-N$, $t+N$.

In order to clarify the entanglement structure, the notion of distributed modes can be introduced~\cite{Alexander.pra2016}. 
Distributed modes, specified by $(L,t)$ with $L\in\{A,B,C,D\}$ and $t\in\mathbb{Z}$, are micronodes obtained by changing the mode decomposition of each macronode from the local micronodes $(\ell,t)$ with $\ell\in\{a,b,c,d\}$ and $t\in\mathbb{Z}$, with the relation 
\begin{align}
\begin{pmatrix}\hat{a}_{A,t}\\ \hat{a}_{B,t}\\ \hat{a}_{C,t}\\ \hat{a}_{D,t}\end{pmatrix} 
=\mathcal{B}^{-1}\begin{pmatrix}\hat{a}_{a,t}\\ \hat{a}_{b,t}\\ \hat{a}_{c,t}\\ \hat{a}_{d,t}\end{pmatrix}. 
\label{eq:distributed_mode}
\end{align}
Using these distributed modes, the nullifiers in Eq.~\eqref{eq:nullifer_QRL_local} become the nullifiers of two-mode squeezed states, 
\begin{align}
\{ & \hat{p}_{A,t}+\hat{p}_{B,t+1}, \hat{x}_{B,t+1}-\hat{x}_{A,t}, \notag\\ 
& \hat{p}_{C,t}+\hat{p}_{D,t+N}, \hat{x}_{D,t+N}-\hat{x}_{C,t}\}_{t\in\mathbb{Z}}. 
\end{align}
Compared with Eq.~\eqref{eq:nullifier_TS_delayed}, we can see the distributed modes correspond to the two-mode squeezed states just before the foursplitter. 
By using the distributed modes and the macronodes, the QRL cluster state is depicted as Fig.~\ref{fig:QRL_diagram}. 
The notion of distributed modes is also important to understand the quantum information flow in MBQC, explained in Sec.~\ref{sec:macronode}.

\section{Macronode operations}
\label{sec:macronode}

\subsection{Description with distributed modes}

\begin{figure}[t]
\centering
\includegraphics[scale=1.0]{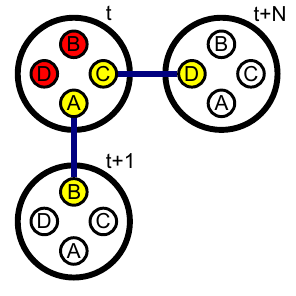}
\caption{Diagram of the situation to consider a single calculation step by consuming the macronode $t$.}
\label{fig:macronode}
\end{figure}

Now we describe each calculation step done by consuming a macronode by measurements. 
The considered situation is depicted in Fig.~\ref{fig:macronode}. 
The distributed modes $(B,t)$ and $(D,t)$ have, instead of halves of two-mode squeezed pairs, input quantum states $\ket{\psi_1}_{B,t}$ and $\ket{\psi_2}_{D,t}$, respectively. 
These input states may be considered as the output of previous calculation steps. 
They will be teleported to $(B,t+1)$ and $(D,t+N)$ by measurements of $(a,t)$, $(b,t)$, $(c,t)$, $(d,t)$ followed by appropriate feedforward via the quantum correlations of the two-mode squeezed vacuum states $\ket{\text{TS}}_{(A,t),(B,t+1)}$ and $\ket{\text{TS}}_{(C,t),(D,t+N)}$. 
(In our notation, for multimode operators or states, we specify the modes by subscripts such as $\ell\ell^\prime,t$ if the time indices coincide, while by subscripts such as $(\ell,t),(\ell^\prime,t^\prime)$ if the time indices are different.)

\begin{figure}[t]
\centering
\includegraphics[scale=0.8]{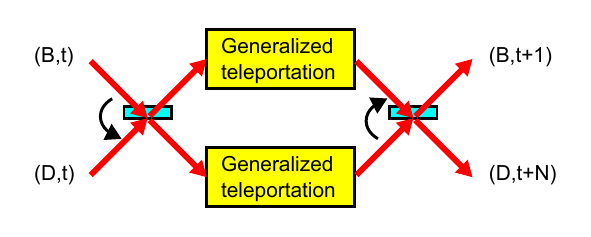}
\caption{Diagram of a single calculation step by consuming the macronode $t$. Quantum states in distributed modes $(B,t)$ and $(D,t)$ are converted by a Mach-Zehnder interferometer with a generalized teleportation in each arm, and output to distributed modes $(B,t+1)$ and $(D,t+N)$.}
\label{fig:macronode_operation}
\end{figure}

The QRL cluster state in this situation is 
\begin{align}
&\ket{\psi_1}_{B,t}\ket{\psi_2}_{D,t}\ket{\text{TS}}_{(A,t),(B,t+1)}\ket{\text{TS}}_{(C,t),(D,t+N)}\ket{\text{rest}} \notag\\
&=  \hat{\mathcal{B}}_{abcd,t}\ket{\psi_1}_{b,t}\ket{\psi_2}_{d,t} \notag\\
&\qquad \ket{\text{TS}}_{(a,t),(B,t+1)}\ket{\text{TS}}_{(c,t),(D,t+N)}\ket{\text{rest}}, 
\end{align}
where the irrelevant qumodes are expressed by $\ket{\text{rest}}$. 
Local measurements are performed on the micronodes $(a,t)$, $(b,t)$, $(c,t)$, $(d,t)$, which we express with some measurement operators $\hat{M}_{a,t}$, $\hat{M}_{b,t}$, $\hat{M}_{c,t}$, $\hat{M}_{d,t}$, respectively. 
We use the symmetry of the foursplitter operation, 
\begin{align}
\hat{\mathcal{B}}_{abcd,t}
&=\hat{B}_{bd,t}\hat{B}_{ac,t}\hat{B}_{cd,t}^\dagger\hat{B}_{ab,t}^\dagger \notag\\
&=\hat{B}_{cd,t}^\dagger\hat{B}_{ab,t}^\dagger\hat{B}_{bd,t}\hat{B}_{ac,t}. 
\end{align}
In addition, we also use a symmetry of two-mode squeezed states,
\begin{align}
\hat{B}_{jl}\ket{\text{TS}}_{jk}\ket{\text{TS}}_{lm} = \hat{B}_{km}^\dagger\ket{\text{TS}}_{jk}\ket{\text{TS}}_{lm}. 
\end{align}
An easy check of this symmetry is done with nullifiers, where the nullifiers calculated by $\hat{U}\hat{\mathcal{N}}\hat{U}^\dagger$ for the left-hand side are expressed as linear combinations of those for the right-hand side. 
Combining them, we finally get, 
\begin{align}
&\hat{M}_{d,t}\hat{M}_{c,t}\hat{M}_{b,t}\hat{M}_{a,t}\hat{\mathcal{B}}_{abcd,t}\ket{\psi_1}_{b,t}\ket{\psi_2}_{d,t} \notag\\
&\qquad \ket{\text{TS}}_{(a,t),(B,t+1)}\ket{\text{TS}}_{(c,t),(D,t+N)}\ket{\text{rest}} \notag\\
&= \hat{B}_{(B,t+1),(D,t+N)}^\dagger(\hat{M}_{d,t}\hat{M}_{c,t}\hat{B}_{cd,t}^\dagger) \notag\\
&\qquad (\hat{M}_{b,t}\hat{M}_{a,t}\hat{B}_{ab,t}^\dagger)(\hat{B}_{bd,t}\ket{\psi_1}_{b,t}\ket{\psi_2}_{d,t}) \notag\\
&\qquad \ket{\text{TS}}_{(a,t),(B,t+1)}\ket{\text{TS}}_{(c,t),(D,t+N)}\ket{\text{rest}}. 
\end{align}
In particular, we consider homodyne measurements since we are here interested in Gaussian operations, and the measured quadrature for each micronode $\hat{p}_j(\theta_j)=\hat{p}_j\cos\theta_j+\hat{x}_j\sin\theta_j$ has a parameter $\theta_j$. 
In this case, $\hat{M}_{b,t}\hat{M}_{a,t}\hat{B}_{ab,t}^\dagger$ and $\hat{M}_{d,t}\hat{M}_{c,t}\hat{B}_{cd,t}^\dagger$ represent generalized Bell measurements, and thus, combined with
 $\ket{\text{TS}}_{(a,t),(B,t+1)}$ and $\ket{\text{TS}}_{(c,t),(D,t+N)}$, they represents generalized quantum teleportation (assuming appropriate feedforward). 
Thus, the single-step operation on $\ket{\psi_1}_{B,t}\ket{\psi_2}_{D,t}$, consuming the macronode $t$, is of the form, 
\begin{align}
&\hat{G}_{BD,t}(\theta_d,\theta_c,\theta_b,\theta_a) \notag\\
&:= \hat{B}_{BD,t}^\dagger \hat{V}_{D,t}(\theta_d, \theta_c)\hat{V}_{B,t}(\theta_b, \theta_a)\hat{B}_{BD,t}, 
\label{eq:macronode_teleportation}
\end{align}
and the quantum states are teleported form $(B,t)$, $(D,t)$ to $(B,t+1)$, $(D,t+N)$. 
Here, $\hat{V}_{B,t}(\theta_b, \theta_a)$ and $\hat{V}_{D,t}(\theta_d, \theta_c)$ are operations induced by the generalized teleportation described in Sec.~\ref{ssec:generalized_teleportation}. 
As shown in Fig.~\ref{fig:macronode_operation}, the single-step transformation $\hat{G}_{BD,t}(\theta_d,\theta_c,\theta_b,\theta_a)$ shares similarity with a Mach-Zehnder interferometer, where a teleportation-based transformation is inserted in each arm.

\subsection{Generalized teleportation}
\label{ssec:generalized_teleportation}

Now we describe the transformation $\hat{V}(\theta_b, \theta_a)$ induced by the generalized teleportation in Heisenberg representation. 
The situation is as follows: 
the input state is in $b$, the two-mode squeezed state is in $a$ and $b^\prime$, and the input state is teleported from $b$ to $b^\prime$ by the generalized Bell measurement $\hat{M}_{b}\hat{M}_{a}\hat{B}_{ab}^\dagger$ on $a$ and $b$, followed by appropriate feedforward to $b^\prime$. 
First, $a$ and $b$ are mixed by the beamsplitter $\hat{B}_{ab}^\dagger$, resulting in, 
\begin{align}
\begin{pmatrix}
\hat{a}_{\tilde{a}} \\ \hat{a}_{\tilde{b}}
\end{pmatrix}
=\hat{B}_{ab}
\begin{pmatrix}
\hat{a}_a \\ \hat{a}_b
\end{pmatrix}
\hat{B}_{ab}^\dagger
=\frac{1}{\sqrt{2}}\begin{pmatrix}1 & 1 \\ -1 & 1\end{pmatrix}\begin{pmatrix}\hat{a}_a \\ \hat{a}_b\end{pmatrix}. 
\end{align}
Then quadratures $\hat{p}_{\tilde{a}}(\theta_a)$ and $\hat{p}_{\tilde{b}}(\theta_b)$ are measured [the definition of rotated quadratures is in Eq.~\eqref{eq:quadrature}], which are, 
\begin{align}
& \sqrt{2}
\begin{pmatrix}
\hat{p}_{\tilde{a}}(\theta_a) \\ \hat{p}_{\tilde{b}}(\theta_b)
\end{pmatrix}
= \sqrt{2}
\begin{pmatrix}
\hat{p}_{\tilde{a}}\cos\theta_a + \hat{x}_{\tilde{a}}\sin\theta_a \\ \hat{p}_{\tilde{b}}\cos\theta_b + \hat{x}_{\tilde{a}}\sin\theta_b
\end{pmatrix} \notag\\
&=
\begin{pmatrix}
\sin\theta_a & -\cos\theta_a \\ -\sin\theta_b & \cos\theta_b
\end{pmatrix}
\begin{pmatrix}
\hat{x}_{a} \\ -\hat{p}_{a}
\end{pmatrix}
+
\begin{pmatrix}
\sin\theta_a & \cos\theta_a \\ \sin\theta_b & \cos\theta_b 
\end{pmatrix}
\begin{pmatrix}
\hat{x}_{b} \\ \hat{p}_{b}
\end{pmatrix}
\label{eq:general_Bell}
\end{align}
On the other hand, the two-mode squeezed state has squeezed quadratures, 
\begin{align}
\begin{pmatrix}
\hat{x}_{b^\prime} - \hat{x}_a \\
\hat{p}_{b^\prime} + \hat{p}_a 
\end{pmatrix}=\hat{\delta}, 
\label{eq:tms}
\end{align}
where $\hat{\delta}$ is the noise term for finite squeezings, and approaches to $0$ in the limit of infinite squeezings.

Eq.~\eqref{eq:tms} is transformed (here omitting $\hat{\delta}$), using Eq.\eqref{eq:general_Bell}, as
\begin{align}
&\begin{pmatrix}
\hat{x}_{b^\prime} \\
\hat{p}_{b^\prime}
\end{pmatrix}=
\begin{pmatrix}
\hat{x}_a \\
-\hat{p}_a 
\end{pmatrix} \notag\\
& = \frac{1}{\sin(\theta_b-\theta_a)}
\begin{pmatrix}
\cos\theta_b & \cos\theta_a \\ \sin\theta_b & \sin\theta_a 
\end{pmatrix}
\begin{pmatrix}
\sin\theta_a & \cos\theta_a \\ \sin\theta_b & \cos\theta_b 
\end{pmatrix}
\begin{pmatrix}
\hat{x}_b \\
\hat{p}_b
\end{pmatrix}\notag\\
& \qquad - \frac{\sqrt{2}}{\sin(\theta_b-\theta_a)}
\begin{pmatrix}
\cos\theta_b & \cos\theta_a \\ \sin\theta_b & \sin\theta_a 
\end{pmatrix}
\begin{pmatrix}
\hat{p}_{\tilde{a}}(\theta_a) \\ \hat{p}_{\tilde{b}}(\theta_b)
\end{pmatrix}
\label{eq:teleportation}. 
\end{align}
Here, the inverse matrix
\begin{align}
&\begin{pmatrix}
\sin\theta_a & -\cos\theta_a \\ -\sin\theta_b & \cos\theta_b
\end{pmatrix}^{-1} \notag\\
& =\frac{1}{\sin(\theta_a-\theta_b)}
\begin{pmatrix}
\cos\theta_b & \cos\theta_a \\ \sin\theta_b & \sin\theta_a 
\end{pmatrix}
\end{align}
is used.  
The last term in Eq.~\eqref{eq:teleportation} can be cancelled by feedforward displacement, because $\hat{p}_{\tilde{a}}(\theta_a)$ and $\hat{p}_{\tilde{b}}(\theta_b)$ are the observables measured in the generalized Bell measurement and $\theta_a$ and $\theta_b$ are known parameters. 
The remaining part represents the generalized teleportation from $b$ to $b^\prime$, accompanied by some transformation determined by the choice of the measurement basis. 
The transformation induced by the generalized teleportation is, 
\begin{align}
& \hat{V}_b^\dagger(\theta_b, \theta_a)
\begin{pmatrix}
\hat{x}_b \\ \hat{p}_b
\end{pmatrix}
\hat{V}_b(\theta_b, \theta_a) \notag\\
& = \frac{1}{\sin(\theta_b-\theta_a)}
\begin{pmatrix}
\cos\theta_b & \cos\theta_a \\ \sin\theta_b & \sin\theta_a 
\end{pmatrix}
\begin{pmatrix}
\sin\theta_a & \cos\theta_a \\ \sin\theta_b & \cos\theta_b 
\end{pmatrix}
\begin{pmatrix}
\hat{x}_b \\ \hat{p}_b
\end{pmatrix} \notag\\
& =: V(\theta_b, \theta_a)
\begin{pmatrix}
\hat{x}_b \\ \hat{p}_b
\end{pmatrix}. 
\label{eq:general_teleportation}
\end{align}
The original quantum teleportation, corresponding to the identity operation, is obtained at $(\theta_b,\theta_a) = (\pi/2,0)$. 
On the other hand, $\theta_b=\theta_a=\theta$ demolishes the input state, because the quadrature $\hat{p}_{b}(\theta)$ of the input state is measured.

The transformation matrix $V(\theta_b, \theta_a)$ is decomposed, by using $\theta_+ :=(\theta_b + \theta_a)/2$ and $\theta_- := (\theta_b - \theta_a)/2$, as 
\begin{align}
& V(\theta_b, \theta_a) \notag\\
& = \begin{pmatrix}
\sin\theta_{+} & \cos\theta_{+} \\
-\cos\theta_{+} & \sin\theta_{+}
\end{pmatrix}
\begin{pmatrix}
\tan\theta_{-} & 0 \\
0 & \cot\theta_{-}
\end{pmatrix}
\begin{pmatrix}
\cos\theta_{+} & -\sin\theta_{+} \\
\sin\theta_{+} & \cos\theta_{+}
\end{pmatrix} \notag\\
& = R\left(\theta_+ -\frac{\pi}{2}\right)S(\tan\theta_-)R(\theta_+), 
\label{eq:general_teleportation_decomposed}
\end{align}
where $R(\theta)$ is a matrix corresponding to a phase-shift operation and $S(\xi)$ is a matrix corresponding to a squeezing operation. 
In particular, a phase-shift operation without squeezing is obtained by setting as $\theta_b = \theta_a+\pi/2$. 
On the other hand, a squeezing operation is obtained depending on $\theta_{-}$; however, most of squeezing operations are accompanied by some phase shifts. 
For example, for $\theta_a=-\theta_b$, the squeezing of the $\hat{x}$ or $\hat{p}$ quadrature is obtained, but it is followed by an additional $-\pi/2$ phase shift. 
If such phase shifts accompanied by the squeezing is inconvenient, a squeezing operation as a one-parameter group is obtained with the squeezing axis in the $\pm\pi/4$ direction, 
\begin{align}
V\left(\theta+\frac{\pi}{2},-\theta\right) 
& = R\left(-\frac{\pi}{4}\right)S(\tan\theta_-)R\left(\frac{\pi}{4}\right) \notag\\
& = 
\begin{pmatrix}
\csc 2\theta_{-} & \cot 2\theta_{-} \\
\cot 2\theta_{-} & \csc 2\theta_{-} 
\end{pmatrix}
\end{align}

An important relation is $S(\tan(-\theta))=-S(\tan\theta)=R(\pi)S(\tan\theta)$, which implies 
\begin{align}
\hat{V}(\theta_b, \theta_a)=\hat{R}(\pi)\hat{V}(\theta_a, \theta_b), 
\label{eq:teleportation_flip}
\end{align}
for arbitrary $\theta_b$ and $\theta_a$.

\subsection{Properties of macronode operations}

Now taking into account the concrete form of the generalized teleportations in Eqs.~\eqref{eq:general_teleportation} and \eqref{eq:general_teleportation_decomposed}, we discuss properties of the whole transformation $\hat{G}_{BD,t}(\theta_d,\theta_c,\theta_b,\theta_a)$ in Eq.~\eqref{eq:macronode_teleportation}.

\subsubsection{Single-mode operation}
\label{sssec:single-mode}

A single-mode operation achievable by $\hat{V}(\theta_b, \theta_a)$ can be simultaneously applied to $(B,t)$ and $(D,t)$, because of commutability between a beamsplitter operation and common single-mode Gaussian operations (without displacements)~\cite{Alexander.pra2016}, 
\begin{align}
\hat{G}_{BD,t}(\theta_b,\theta_a,\theta_b,\theta_a)
&= \hat{B}_{BD,t}^\dagger \hat{V}_{D,t}(\theta_b, \theta_a)\hat{V}_{B,t}(\theta_b, \theta_a)\hat{B}_{BD,t} \notag\\
&= \hat{B}_{BD,t}^\dagger \hat{B}_{BD,t} \hat{V}_{D,t}(\theta_b, \theta_a)\hat{V}_{B,t}(\theta_b, \theta_a) \notag\\
&= \hat{V}_{D,t}(\theta_b, \theta_a)\hat{V}_{B,t}(\theta_b, \theta_a). 
\end{align}
The commutability comes from the fact that the transformation matrix of a beamsplitter is a real matrix in Eq.~\eqref{eq:beamsplitter}, which means that the matrices for annihilation and creation operators coincide.

This property can be used to implement a single-mode operation $\hat{V}(\theta_b, \theta_a)$, by setting an input state only in $(B,t)$ or $(D,t)$. 
It is known that two steps of generalized teleportation is enough to implement an arbitrary single-mode Gaussian unitary operation~\cite{Ukai.pra2010,Alexander.pra2014}. 
Among single-mode operations, phase-shift operations $\hat{R}(2\theta)$ obtained by $(\theta_d,\theta_c,\theta_b,\theta_a)=(\theta+\pi/2,\theta,\theta+\pi/2,\theta)$ is used in the decomposition of a beamsplitter network in Sec.~\ref{sec:beamsplitter_network}.

\subsubsection{Swapping of outputs}

The input states in $(B,t)$ and $(D,t)$ are teleported to $(B,t+1)$, $(D,t+N)$ after the conversion by $\hat{G}_{BD,t}(\theta_d,\theta_c,\theta_b,\theta_a)$; however, the destinations can be swapped by swapping $\theta_b$ and $\theta_a$. 
Eq.~\eqref{eq:teleportation_flip} shows that the swapping of measurement phases result in a $\pi$ phase shift, which acts as a sign flip of the annihilation operator. 
The effect of a sign flip of $\hat{a}_{B,t}$ for the output of the final beamsplitter $\hat{B}_{BD,t}^\dagger$ is, 
\begin{align}
\hat{B}_{BD,t} 
\begin{pmatrix}
\pm \hat{a}_{B,t} \\ \hat{a}_{D,t}
\end{pmatrix}
\hat{B}_{BD,t}^\dagger
& = 
\frac{1}{\sqrt{2}}\begin{pmatrix}1 & 1 \\ -1 & 1\end{pmatrix} 
\begin{pmatrix}
\pm \hat{a}_{B,t} \\ \hat{a}_{D,t}
\end{pmatrix} \notag\\
& = \frac{1}{\sqrt{2}}\begin{pmatrix}\pm \hat{a}_{B,t} + \hat{a}_{D,t} \\ \mp \hat{a}_{B,t} + \hat{a}_{D,t}\end{pmatrix}. 
\end{align}
Apparently the two output is swapped by the sign flip of $\hat{a}_{B,t}$. 
Using the swapping property, the wires to transfer the input states can be flexibly designed on the QRL cluster state.

\subsubsection{Beamsplitter interaction}

As an example of two-mode operations, beamsplitter interactions are fundamentally important. 
They are photon-number conserving operations, and thus obtained by restricting $\hat{V}(\theta_b, \theta_a)$ to photon-number conserving operations, that is, phase shifts. 
By setting $(\theta_d,\theta_c,\theta_b,\theta_a)=(\theta_2+\pi/2,\theta_2,\theta_1+\pi/2,\theta_1)$, we obtain, 
\begin{align}
& \hat{G}_{BD,t}^\dagger(\theta_d,\theta_c,\theta_b,\theta_a)
\begin{pmatrix}
\hat{a}_{B,t} \\ \hat{a}_{D,t}
\end{pmatrix}
\hat{G}_{BD,t}(\theta_d,\theta_c,\theta_b,\theta_a) \notag\\
& = \frac{1}{2}
\begin{pmatrix}
1 & 1 \\
-1 & 1
\end{pmatrix}
\begin{pmatrix}
e^{2i\theta_1} & 0 \\
0 & e^{2i\theta_2}
\end{pmatrix}
\begin{pmatrix}
1 & -1 \\
1 & 1
\end{pmatrix}
\begin{pmatrix}
\hat{a}_{B,t} \\ \hat{a}_{D,t}
\end{pmatrix} \notag\\
& = e^{i(\theta_2+\theta_1)}
\begin{pmatrix}
\cos(\theta_2-\theta_1) & i\sin(\theta_2-\theta_1) \\
i\sin(\theta_2-\theta_1) & \cos(\theta_2-\theta_1)
\end{pmatrix}
\end{align}
Therefore, the beamsplitter operation achievable with a single macromode is of the form, 
\begin{align}
\hat{T}_{jk}(\tau,\phi) & = \hat{R}_j(\phi)\hat{R}_k(\phi)\hat{B}_{jk}^\prime(\tau), 
\label{eq:macronode_beamsplitter}
\end{align}
where $\tau=\theta_2-\theta_1$, $\phi=\theta_2+\theta_1$ and $\hat{B}_{jk}^\prime(\tau)=\hat{R}_k(\pi/2)\hat{B}_{jk}(\tau)\hat{R}_k(-\pi/2)$. 
Thus, $\hat{T}_{jk}(\tau,\phi)$ is a beamsplitter with a tunable reflectivity $\hat{B}_{jk}^\prime(\tau)$ followed by a tunable phase shift $\hat{R}_j(\phi)\hat{R}_k(\phi)$ which is common to both output ports. 
(Here we consider the common phase at the output for the sake of later convenience, though actually the common phase can be instead at the input, because the common phase commutes with a beamsplitter.)

In Sec.~\ref{sec:beamsplitter_network}, we will decompose a beamsplitter network with this $\hat{T}_{jk}(\tau,\phi)$.

\section{Decomposition of beamsplitter networks}
\label{sec:beamsplitter_network}

The conversion of $N$ qumodes by an arbitrary beamsplitter network $\hat{U}$ is expressed by a $N\times N$ unitary matrix $U(N)$ as in Eq.~\eqref{eq:beamsplitter_network}. 
In Ref.~\cite{Reck.prl1994} by Reck  \textit{et al.}, an efficient decomposition of a general beamsplitter network with beamsplitters and phase shifts is given. 
However, as explained below, the components of the decomposition was $\hat{C}_{jk}(\tau,\phi) = \hat{B}_{jk}^\prime(\tau)\hat{R}_j(\phi)$ (or its variant). 
Therefore, we cannot use the decomposition directly, since the components available with a single macronode operation are $\hat{T}_{jk}(\tau,\phi) = \hat{R}_j(\phi)\hat{R}_k(\phi)\hat{B}_{jk}^\prime(\tau)$ in Eq.~\eqref{eq:macronode_beamsplitter}. 
In the following, we will omit the prime for the beamsplitter operator, and also omit the arguments $(\tau,\phi)$ when they are not important.

In order to decompose an arbitrary beamsplitter network into macronode operations, here we take the following three steps. 
First, we start from the decomposition of a beamsplitter network $\hat{U}$ into $\hat{C}_{jk}=\hat{B}_{jk}\hat{R}_j$ (beamsplitter with a single input phase shift) and residual $\hat{R}_j$. 
Next, we reinterpret it as the decomposition into $\hat{S}_{jk}=\hat{R}_j\hat{B}_{jk}$ (beamsplitter with a single output phase shift) and residual $\hat{R}_j$. 
Finally, we convert it to the decomposition into $\hat{T}_{jk} = \hat{R}_j\hat{R}_k\hat{B}_{jk}$ (beamsplitter with a common output phase shift) and residual $\hat{R}_j$.

In order to decompose $\hat{U}$ with $\hat{C}_{jk}$, we consider the matrix expression of $\hat{C}_{jk}^{-1}=\hat{R}_j^{-1}\hat{B}_{jk}^{-1}$. 
\begin{align}
& [\hat{C}_{jk}^{-1}(\tau,\phi)]^\dagger
\begin{pmatrix}
\hat{a}_j \\ \hat{a}_k
\end{pmatrix}
\hat{C}_{jk}^{-1}(\tau,\phi) \notag\\
& = 
\begin{pmatrix}
e^{-i\phi} & 0 \\
0 & 1
\end{pmatrix}
\begin{pmatrix}
\cos\tau & -i\sin\tau \\
-i\sin\tau & \cos\tau
\end{pmatrix}
\begin{pmatrix}
\hat{a}_j \\ \hat{a}_k
\end{pmatrix} \notag\\
& = 
\begin{pmatrix}
e^{-i\phi}\cos\tau & -ie^{-i\phi}\sin\tau \\
-i\sin\tau & \cos\tau
\end{pmatrix}
\begin{pmatrix}
\hat{a}_j \\ \hat{a}_k
\end{pmatrix} \notag\\
& := 
\begin{pmatrix}
a(\tau,\phi) & b(\tau,\phi) \\
c(\tau) & d(\tau)
\end{pmatrix}
\begin{pmatrix}
\hat{a}_j \\ \hat{a}_k
\end{pmatrix}. 
\end{align}

\begin{figure}[t]
\centering
\includegraphics[scale=0.8]{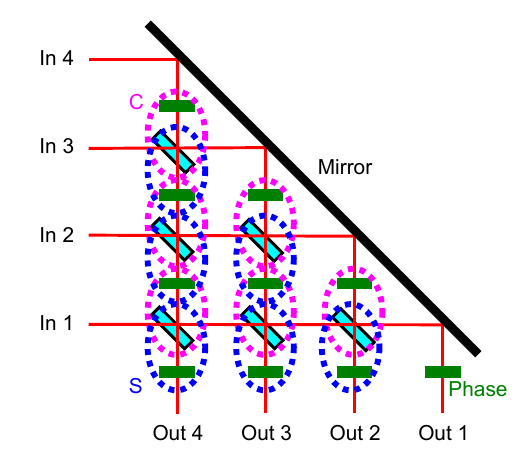}
\caption{Beamsplitter network decomposed into beamsplitters and phases for $N=4$. The two different pairings of a beamsplitter and a phase, $\hat{C}_{jk}$ and $\hat{S}_{jk}$, are shown by dotted circles.}
\label{fig:beamsplitter_network_decomp}
\end{figure}

The decomposition in Ref.~\cite{Reck.prl1994} is as follows. 
We consider multiplying $C_{N,N-1}^{-1}(\tau_{N,N-1},\phi_{N,N-1})$ to $U(N)$ from the right. 
Here $C_{N,N-1}^{-1}$ is the $N\times N$ matrix corresponding to $\hat{C}_{N,N-1}^{-1}$. 
\begin{align}
U(N)C_{N,N-1}^{-1}
=
\begin{pmatrix}
u_{11} & \dots &u_{1,N}\\
\vdots & \ddots & \vdots \\
u_{N,1} & \dots  & u_{N,N}
\end{pmatrix}
\begin{pmatrix}
I(N-2) & 0 & 0 \\
0 & d & c \\
0 & b & a
\end{pmatrix}.
\end{align}
The $(N-1,N)$-element of $U(N)C_{N,N-1}^{-1}$ is, 
\begin{align}
& (U(N)C_{N,N-1}^{-1})_{N,N-1} \notag\\
& = u_{N-1,N-1}c(\tau_{N,N-1})+u_{N-1,N}a(\tau_{N,N-1},\phi_{N,N-1})
\end{align}
For arbitrary $u_{N-1,N-1}$ and $u_{N-1,N}$ with complex values, there is a choice of the parameters $(\tau_{N,N-1},\phi_{N,N-1})$ that cancels the $(N-1,N)$-element of $U(N)C_{N,N-1}^{-1}$. 
(The absolute values of the first and second terms are tuned to be equal via $\tau_{N,N-1}$, and then they are cancelled by adjusting the argument via $\phi_{N,N-1}$.)
We can see that we cannot directly use $\hat{T}_{N,N-1}$ or $\hat{T}_{N,N-1}^{-1}$ for this cancellation because the phase factor is common for both modes. 
This is why we first have to consider $\hat{C}_{jk}$ instead of $\hat{T}_{j,k}$. 
Repeating this from $C_{N,N-1}^{-1}$ to $C_{N,1}^{-1}$, the off-diagonal elements from $(N,N-1)$ to $(N,1)$ are cancelled. 
Since $U(N)$ is unitary, after these cancellations, the off-diagonal elements from $(N-1,N)$ to $(1,N)$ are also zero. 
\begin{align}
U(N)C_{N,N-1}^{-1}\dots C_{N,1}^{-1}
=
\begin{pmatrix}
U(N-1) & 0 \\
0 & e^{i\phi_{N,0}}
\end{pmatrix}. 
\end{align}
Now the remaining part is a $(N-1)\times(N-1)$ unitary matrix $U(N-1)$. 
We further recursively repeat the cancellations to decrease the size of the unitary matrix, and finally obtain a diagonal matrix, 
\begin{align}
& U(N)(C_{N,N-1}^{-1}\dots C_{N,1}^{-1})(C_{N-1,N-2}^{-1}\dots C_{N-1,1}^{-1})\dots C_{2,1}^{-1} \notag\\
& \qquad = 
\begin{pmatrix}
e^{i\phi_{1,0}} &  & 0 \\
 & \ddots & \\
0 & & e^{i\phi_{N,0}}
\end{pmatrix} 
:= D. 
\end{align}
This means, 
\begin{align}
U(N)=DC_{2,1}C_{3,1}C_{3,2}\dots C_{N,N-2}\hat{C}_{N,N-1}, 
\end{align}
\begin{align}
\hat{U}
&=\hat{R}_1\dots\hat{R}_N\hat{C}_{2,1}\hat{C}_{3,1}\hat{C}_{3,2}\dots\hat{C}_{N,N-2}\hat{C}_{N,N-1} \notag\\
&=\hat{R}_1(\hat{R}_2\hat{C}_{2,1})(\hat{R}_3\hat{C}_{3,1}\hat{C}_{3,2})\dots(\hat{R}_N\hat{C}_{N,1}\dots\hat{C}_{N,N-1}). 
\end{align}
Explicitly showing the parameters, now we obtained the decomposition into $\hat{C}_{j,k}(\tau_{j,k},\phi_{j,k})=\hat{B}_{j,k}(\tau_{j,k})\hat{R}_{j}(\phi_{j,k})$ and $\hat{R}_j(\phi_{j,0})$.

Next, we switch the pairing of a beamsplitter and a phase shift from $\hat{C}_{jk}=\hat{B}_{jk}\hat{R}_j$ to $\hat{S}_{jk}=\hat{R}_j\hat{B}_{jk}$, as 
\begin{align}
\hat{R}_j(\phi_{j,k-1})\hat{C}_{jk}(\tau_{j,k},\phi_{j,k}) = \hat{S}_{jk}(\tau_{j,k},\phi_{j,k}-1)\hat{R}_j(\phi_{j,k})
\end{align}
This rearrangement of the beamsplitter-phase pairs is depicted in Fig.~\ref{fig:beamsplitter_network_decomp} and shown by equations as
\begin{widetext}
\begin{subequations}
\begin{align}
\hat{R}_2(\phi_{2,0})\hat{C}_{2,1}(\tau_{2,1},\phi_{2,1})& =\hat{S}_{2,1}(\tau_{2,1},\phi_{2,0})\hat{R}_2(\phi_{2,1}), \\
\hat{R}_3(\phi_{3,0})\hat{C}_{3,1}(\tau_{3,1},\phi_{3,1})\hat{C}_{3,2}(\tau_{3,2},\phi_{3,2}) 
&=\hat{S}_{3,1}(\tau_{3,1},\phi_{3,0})\hat{S}_{3,2}(\tau_{3,2},\phi_{3,1})\hat{R}_3(\phi_{3,2}), \\
& \qquad\vdots \notag\\
\hat{R}_N(\phi_{N,0})\hat{C}_{N,1}(\tau_{N,1},\phi_{N,1})\dots\hat{C}_{N,N-1}(\tau_{N,N-1},\phi_{N,N-1}) 
&=\hat{S}_{N,1}(\tau_{N,1},\phi_{N,0})\dots\hat{S}_{N,N-1}(\tau_{N,N-1},\phi_{N,N-2})\hat{R}_N(\phi_{N,N-1}). 
\end{align}
\end{subequations}
\end{widetext}
As a result, we get the decomposition
\begin{align}
\hat{U}=\hat{R}_1(\hat{S}_{2,1}\hat{R}_2)(\hat{S}_{3,1}\hat{S}_{3,2}\hat{R}_3)\dots(\hat{S}_{N,1}\dots\hat{S}_{N,N-1}\hat{R}_N). 
\label{eq:decomposition_S}
\end{align}

\begin{figure}[t]
\centering
\includegraphics[scale=0.8]{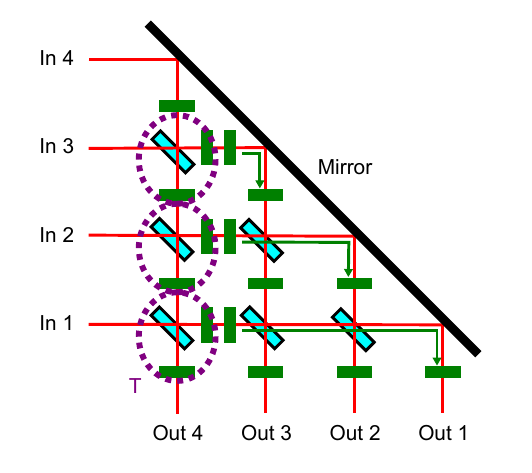}
\caption{Diagram of the phase compensation when converting $\hat{S}_{jk}$ to $\hat{T}_{jk}$, for $\hat{S}_{43}$, $\hat{S}_{42}$, $\hat{S}_{41}$. The compensation phases are absorbed into the phase shifts at the mirror reflection.}
\label{fig:beamsplitter_phase_compensation}
\end{figure}

\begin{figure}[tb]
\centering
\includegraphics[scale=0.7]{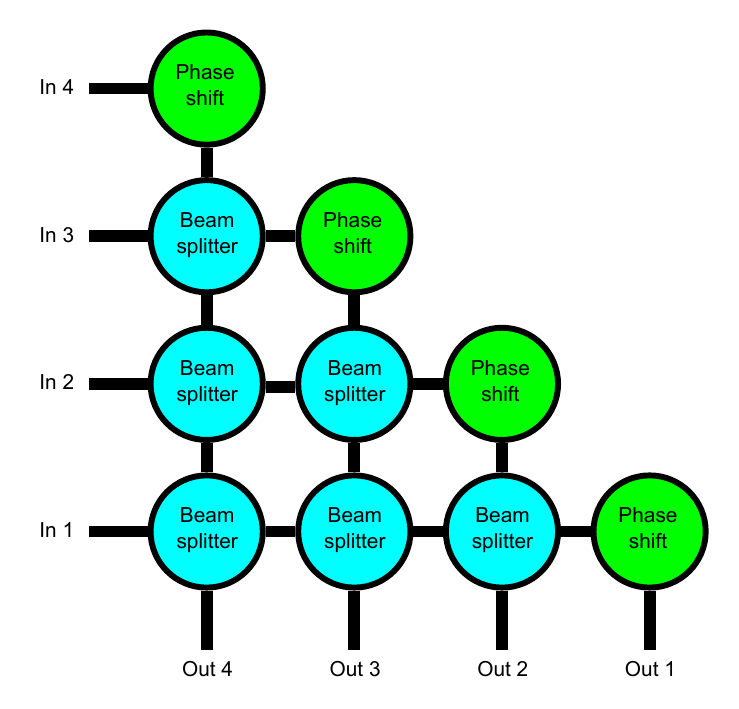}
\caption{Implementation of a general beamsplitter network $U(N)$ on the QRL cluster state. 
Nodes represent macronodes of the QRL cluster state. 
Although the input states come from left in the figure for the convenience of visualization, actually they can start at the micronodes of the first macronodes.}
\label{fig:QRL_beamsplitter_network}
\end{figure}

Finally, we convert the decomposition into $\hat{S}_{jk}=\hat{R}_j\hat{B}_{jk}$ to the decomposition into $\hat{T}_{jk}=\hat{R}_j\hat{R}_k\hat{B}_{jk}$. 
The components $\hat{S}_{jk}$ in Eq.~\eqref{eq:decomposition_S} are converted one by one from right to left, starting from $\hat{S}_{N,N-1}$. 
Each conversion generates an additional compensation phase $\hat{R}_k^{(c)}$, from the relation $\hat{S}_{jk}=\hat{R}_k^{(c)}\hat{T}_{jk}$, where the superscript $(c)$ is used to distinguish the compensation phase from the phase shifts constituting the decomposition in Eq.~\eqref{eq:decomposition_S}. 
We consider absorption of $\hat{R}_k^{(c)}$ into $\hat{R}_k$ by moving $\hat{R}_k^{(c)}$ to the left. 
For example, when $\hat{S}_{N,N-1}$ is converted to $\hat{T}_{N,N-1}$, the compensation phase $\hat{R}_{N-1}^{(c)}$ commute with $\hat{S}_{N,1}\dots\hat{S}_{N,N-2}$ and absorbed into $\hat{R}_{N-1}$. 
However, when $\hat{S}_{N,N-2}$ is converted to $\hat{T}_{N,N-2}$, the compensation phase $\hat{R}_{N-2}^{(c)}$ does not commute with $\hat{S}_{N-1,N-2}$ which exists before $\hat{R}_{N-2}$. 
Thus we must consider the relation of a beamsplitter and phase shifts, 
\begin{align}
\hat{B}_{jk}(\tau)\hat{R}_k(\phi)=\hat{R}_k(\phi)\hat{R}_j(\phi)\hat{B}_{jk}(\tau)\hat{R}_j(-\phi), 
\end{align}
which leads to the parameter change shown below when $\hat{R}_{N-2}^{(c)}$ goes through $\hat{S}_{N-1,N-2}$, 
\begin{align}
& \hat{S}_{N-1,N-2}(\tau,\phi)\hat{R}_{N-1}(\phi^\prime)\hat{R}_{N-2}^{(C)}(\phi^{\prime\prime}) \notag\\
& = \hat{R}_{N-2}^{(C)}(\phi^{\prime\prime})\hat{S}_{N-1,N-2}(\tau,\phi+\phi^{\prime\prime})\hat{R}_{N-1}(\phi^\prime-\phi^{\prime\prime}). 
\end{align}
In general, when $\hat{S}_{j,k}$ is converted to $\hat{T}_{j,k}$, the compensation phase $\hat{R}_k^{(c)}$ goes through $j-k-1$ noncommutative terms $\hat{S}_{k+1,k}$, $\dots$, $\hat{S}_{j-1,k}$ before it is absorbed into $\hat{R}_k$. 
At these noncommutative terms, the parameters are changed as, 
\begin{subequations}
\begin{align}
& \hat{S}_{k+m,k}(\tau,\phi)\hat{S}_{k+m,k+1}(\tau^\prime,\phi^\prime)\hat{R}_{k}^{(C)}(\phi^{\prime\prime}) \notag\\
& = \hat{R}_{k}^{(C)}(\phi^{\prime\prime})\hat{S}_{k+m,k}(\tau,\phi+\phi^{\prime\prime})\hat{S}_{k+m,k+1}(\tau^\prime,\phi^\prime-\phi^{\prime\prime}), \notag\\
& \qquad \text{for $2 \le m \le j-k-1$}, \\
& \hat{S}_{k+1,k}(\tau,\phi)\hat{R}_{k+1}(\phi^\prime)\hat{R}_{k}^{(C)}(\phi^{\prime\prime}) \notag\\
& = \hat{R}_{k}^{(C)}(\phi^{\prime\prime})\hat{S}_{k+1,k}(\tau,\phi+\phi^{\prime\prime})\hat{R}_{k+1}(\phi^\prime-\phi^{\prime\prime}). 
\end{align}
\end{subequations}
The paths of the compensation phases are depicted in Fig.~\ref{fig:beamsplitter_phase_compensation}. 
By repeating the above compensation one by one from $\hat{S}_{N,N-1}$ to $\hat{S}_{2,1}$, we finally obtain the decomposition of a beamsplitter network into macronode operations, 
\begin{align}
\hat{U}=\hat{R}_1(\hat{T}_{2,1}\hat{R}_2)(\hat{T}_{3,1}\hat{T}_{3,2}\hat{R}_3)\dots(\hat{T}_{N,1}\dots\hat{T}_{N,N-1}\hat{R}_N). 
\end{align}
This configuration on the QRL cluster state is depicted in Fig.~\ref{fig:QRL_beamsplitter_network}, which has a triangular shape. 
Since the decomposition in Ref.~\cite{Reck.prl1994} is efficient in the sense there is no redundancy in the degrees of freedom, the above decomposition is also expected to be optimum with respect to the size on the QRL cluster.

\subsection{Another decomposition}

Another efficient decomposition of a general beamsplitter network has been shown by Clements \textit{et al.}\ in Ref.~\cite{Clements.optica2016}, where the beamsplitters are arranged in a rectangular shape. 
The decomposition of an $N$-mode beamsplitter network by Clements \textit{et al.}\ is composed of $N$ layers of beamsplitters $\hat{B}_{j,j+1}^{(l)}$ and associated phase shifts, where the superscript $(l)$ denotes the $l$-th layer, and the mode number $j$ takes only odd or even numbers depending on whether $l$ is odd or even. 
The rectangle-shaped decomposition by Clements \textit{et al.}\ and the triangle-shaped decomposition by Reck \textit{et al.}\ in Ref.~\cite{Reck.prl1994} are equally optimum in the sense that there is no redundancy: the number of beamsplitters for an $N$-mode beamsplitter network is the same, $N(N-1)/2$ for both decompositions.

\begin{figure}[tb]
\centering
\includegraphics[scale=0.5]{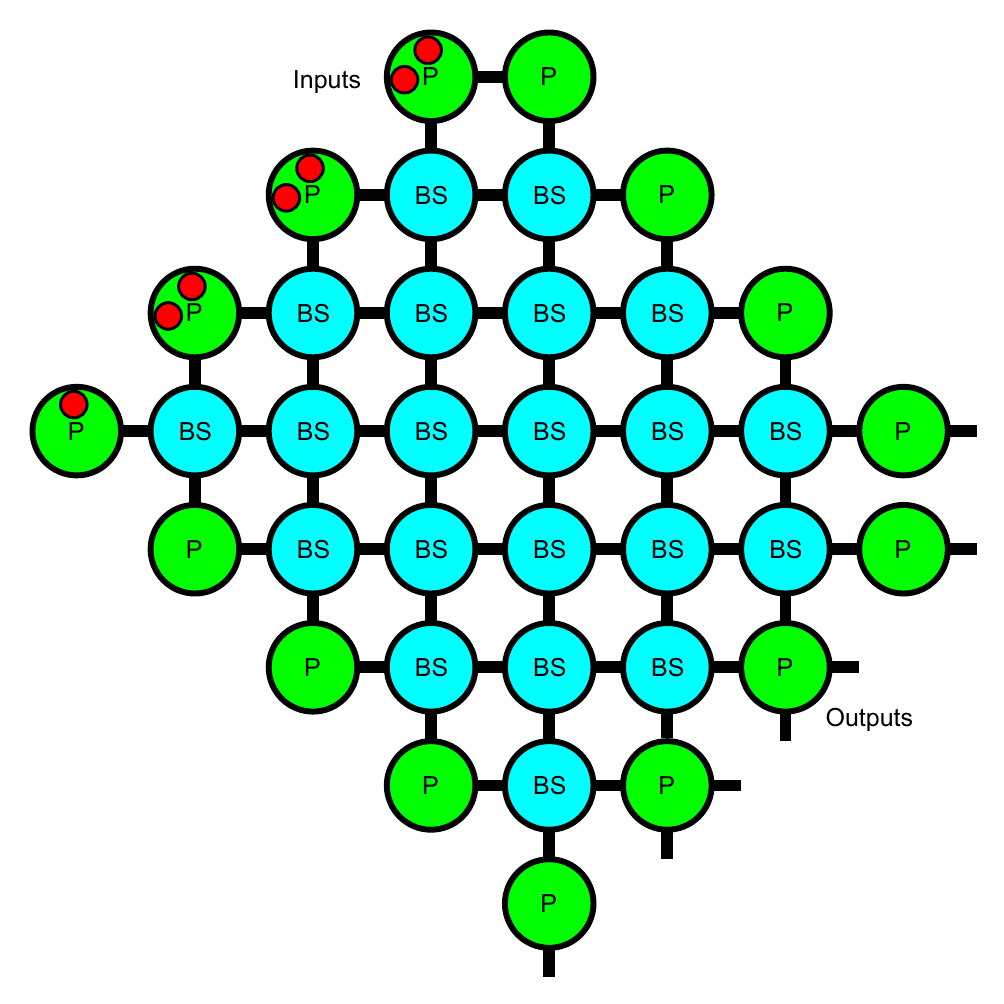}
\caption{Implementation of a general beamsplitter network $U(N)$ on the QRL cluster state in another way, for $N=7$. 
P: phase shift, BS: beamsplitter. 
The beamsplitter nodes have a common output phase as in Eq.~\eqref{eq:macronode_beamsplitter}. 
Phase-shift nodes with two inputs give common phase shifts, as discussed in Sec.~\ref{sssec:single-mode}.}
\label{fig:BS_network_symmetric}
\end{figure}

We propose the configuration on the QRL cluster state shown in Fig.~\ref{fig:BS_network_symmetric}, which corresponds to the rectangle-shaped decomposition by Clements \textit{et al.}\ of a general beamsplitter network. 
Beamsplitter nodes are arranged in the oblique rectangular area, and surrounded by phase-shift nodes. 
Input modes are aligned obliquely, expressed by micronodes in the upper left macronodes, while output modes are expressed by open edges from lower right macronodes. 
Thanks to this oblique alignment, the beamsplitter nodes are efficiently arranged, in contrast to previous proposal~\cite{Alexander.prl2017}, where input modes are aligned vertically and redundant phase shifts are inserted between beamsplitters.

We can validate the configuration in Fig.~\ref{fig:BS_network_symmetric} by checking that the common phase shifts in Eq.~\eqref{eq:macronode_beamsplitter} at the output of the beamsplitter nodes (``BS'' in Fig.~\ref{fig:BS_network_symmetric}) in the $l$-th layer, together with additional phase shifts at turning macronodes (``P'' in Fig.~\ref{fig:BS_network_symmetric}), have sufficient degrees of freedom to realize arbitrary relative phases for input pairs of all beamsplitters in the $(l+1)$-th layer. 
Note that there are phase-shift nodes with two inputs in the input and output layer, which give common phase shifts as discussed in Sec.~\ref{sssec:single-mode}. 
In fact, Fig.~\ref{fig:BS_network_symmetric} has small redundancies in phase shifts, unlike Fig.~\ref{fig:QRL_beamsplitter_network}. 
Some phase shifts can be eliminated, but they are left for visual understandability (phase-shift macronodes surround beamsplitter macronodes).

The advantage of the rectangular decomposition, compared with the triangular one, is uniformity: the difference of the number of beamsplitters among various paths is much smaller.
This uniformity contribute to symmetric losses for finite-squeezing situations by equalizing the number of teleportations~\cite{Alexander.prl2017}, which is often advantageous, depending on applications such as boson samplings~\cite{Aaronson.toc2013}.

On the other hand, compared with the triangular configuration in Fig.~\ref{fig:QRL_beamsplitter_network}, the rectangular configuration in Fig.~\ref{fig:BS_network_symmetric} has densely arranged input modes, which may be disadvantageous in some situations. 
For the triangular case, we can assign one macronode for each input, while for the rectangular case, there are macronodes with two inputs. 
Therefore, for example, for the triangular case, we may use projective measurements on neighboring macronodes to prepare input states via quantum correlation, while if we do the same thing for the rectangular case, the projective measurements become common for two input states. 
Another point is that the connectivity of the beamsplitter network with prior or subsequent processing units may not be good due to the dense arrangement. 
As we will see in Sec.~\ref{sec:Gaussian}, the triangular beamsplitter network is naturally applied to the Bloch-Messiah reduction of general Gaussian unitaries, while the dense input arrangement in the rectangular beamsplitter network may require additional structures as adapters.

\section{Decomposition of Gaussian operations}
\label{sec:Gaussian}

\begin{figure}[t]
\centering
\includegraphics[scale=0.8]{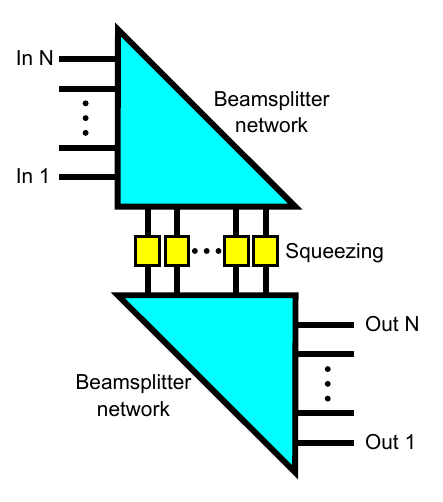}
\caption{Implementation of a general multimode Gaussian operation on the QRL cluster state, based on the Bloch-Messiah reduction. 
Squeezing operations on individual modes, aligned in a line, are sandwiched by beamsplitter networks with the triangular configuration.}
\label{fig:QRL_Gaussian}
\end{figure}

By combining the decomposition of beamsplitter networks in Sec.~\ref{sec:beamsplitter_network} with the Bloch-Messiah reduction of Gaussian unitaries in Eq.~\eqref{eq:Bloch_Messiah}, we can see that an arbitrary multimode Gaussian unitary can be implemented on the QRL cluster state with the configuration that single-mode squeezing macronodes on a line are sandwiched by the two triangle-shaped beamsplitter networks. 
This configuration is depicted in Fig.~\ref{fig:QRL_Gaussian}. 
Note that the phase shifts accompanied by squeezing in Eq.~\eqref{eq:general_teleportation_decomposed} can be absorbed into the former and/or the latter beamsplitter networks. 
We also note that this configuration has good compatibility with cylindrically structured two-dimensional cluster states obtained with the time-domain multiplexing method~\cite{Alexander.pra2016,Asavanant.science2019}.

However, we do not claim the efficiency of the configuration in Fig.~\ref{fig:QRL_Gaussian}. 
We think it is rather guarantee of the maximum size required to implement a general multimode Gaussian operation. 
In contrast to the case of beamsplitter networks where it is natural to restrict the macronode operations to beamsplitter interactions (because the total photon number must be conserved), there is no reason for general Gaussian unitaries to limit the macronode operations to beamsplitter interactions. 
Therefore, we may find more efficient configurations by using general transformations in Eq.~\eqref{eq:macronode_teleportation} instead of beamsplitter interactions. 
This remains an open question for future research. 
At least, we may find subgroups of the Gaussian unitaries which can be implemented with a size similar to the beamsplitter network. 
We find that what we call multimode shear operations are an example of such a subgroup.

A single-mode shear operation is an operation that does not change one of the quadratures while the invariant quadrature is added to the conjugate quadrature, which looks like a shearing of the phase space. 
Mathematically, the $\hat{x}$-invariant shear operation $\hat{H}_x(\kappa):=e^{i\kappa \hat{x}^2/2}$ and the $\hat{p}$-invariant shear operation $\hat{H}_p(\kappa):=e^{i\kappa \hat{p}^2/2}$ is expressed as, 
\begin{subequations}
\begin{align}
\hat{H}_x^\dagger(\kappa)
\begin{pmatrix}
\hat{x} \\ \hat{p}
\end{pmatrix}
\hat{H}_x(\kappa)
&=
\begin{pmatrix}
1 & 0 \\ \kappa & 1
\end{pmatrix}
\begin{pmatrix}
\hat{x} \\ \hat{p}
\end{pmatrix}
\end{align} \\
\begin{align}
\hat{H}_p^\dagger(\kappa)
\begin{pmatrix}
\hat{x} \\ \hat{p}
\end{pmatrix}
\hat{H}_p(\kappa)
&=
\begin{pmatrix}
1 & -\kappa \\ 0 & 1
\end{pmatrix}
\begin{pmatrix}
\hat{x} \\ \hat{p}
\end{pmatrix}. 
\end{align}
\end{subequations}
In particular, the $\hat{x}$-invariant shear gate is also called as a quadratic phase gate~\cite{Miwa.pra2009}, since the unitary transformation works as adding a quadratic phase factor to the wave function, $\hat{H}_x(\kappa)\int dx\,\psi(x)\ket{x} = \int dx\,e^{i\kappa x^2/2}\psi(x)\ket{x}$. 
In fact, these shear operations are available as a macronode operation. 
Taking $(\theta_b,\theta_a)=(\pi/2,\theta)$ or $(\theta+\pi/2,0)$ in Eq.~\eqref{eq:general_teleportation} of the generalized teleportation, we obtain, 
\begin{subequations}
\begin{align}
\hat{V}_b^\dagger\left(\frac{\pi}{2}, \theta\right)
\begin{pmatrix}
\hat{x}_b \\ \hat{p}_b
\end{pmatrix}
\hat{V}_b\left(\frac{\pi}{2}, \theta\right)
& = 
\begin{pmatrix}
1 & 0 \\ 2\tan\theta & 1 
\end{pmatrix}
\begin{pmatrix}
\hat{x}_b \\ \hat{p}_b
\end{pmatrix}. \\
\hat{V}_b^\dagger\left(\theta+\frac{\pi}{2}, 0\right)
\begin{pmatrix}
\hat{x}_b \\ \hat{p}_b
\end{pmatrix}
\hat{V}_b\left(\theta+\frac{\pi}{2}, 0\right)
& = 
\begin{pmatrix}
1 & -2\tan\theta \\ 0 & 1 
\end{pmatrix}
\begin{pmatrix}
\hat{x}_b \\ \hat{p}_b
\end{pmatrix}. 
\end{align}
\end{subequations}
Since the quadratures are symmetric, in the following we only consider the $\hat{x}$-invariant case and omit the subscript $x$.

\begin{figure}[t]
\centering
\includegraphics[scale=0.7]{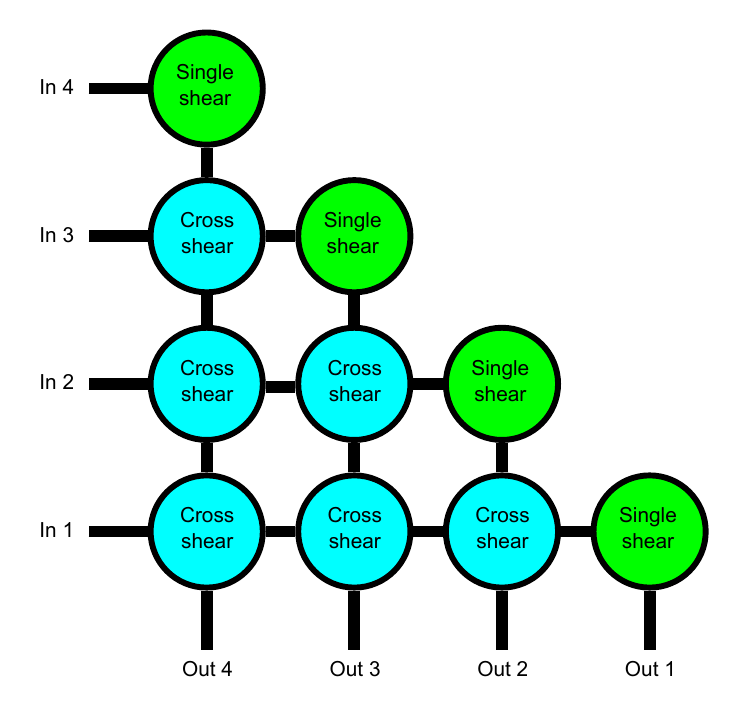}
\caption{Implementation of a general multimode shear operation on the QRL cluster state.
Though the input states come from left in the figure for the convenience of visualization, they can start at the micronodes of the first macronodes.}
\label{fig:QRL_multishear}
\end{figure}

Extending the above single-mode shear operations, $N$-mode shear operations can take cross terms in their quadratic Hamiltonian as, 
\begin{align}
\hat{H}(K):=\exp\left(\frac{i}{2}\sum_{j,k\in\{1,\dots,N\}} K_{jk} \hat{x}_j\hat{x}_k\right), 
\end{align}
where $K$ is an $N\times N$ real symmetric matrix. 
These shear operations are mutually commutative and operator multiplication is compatible with matrix addition $\hat{H}(K)\hat{H}(K^\prime)=\hat{H}(K+K^\prime)$.

We consider two-mode shear operations available as a macronode operation, and use them to construct general $N$-mode shear operations. 
The two-mode shear operations as a macronode operation are obtained by restricting the two generalized teleportations to single-mode shear operations as, 
\begin{align}
& \hat{G}_{BD,t}^\dagger\left(\frac{\pi}{2},\theta_c,\frac{\pi}{2},\theta_a\right)
\begin{pmatrix}
\hat{x}_B \\ \hat{x}_D \\ \hat{p}_B \\ \hat{p}_D
\end{pmatrix}
\hat{G}_{BD,t}\left(\frac{\pi}{2},\theta_c,\frac{\pi}{2},\theta_a\right) \notag\\
& = \frac{1}{2}
\begin{pmatrix}
1 & 1 & 0 & 0 \\
-1 & 1 & 0 & 0 \\
0 & 0 & 1 & 1 \\
0 & 0 & -1 & 1 \\
\end{pmatrix}
\begin{pmatrix}
1 & 0 & 0 & 0 \\
0 & 1 & 0 & 0 \\
2\tan\theta_a & 0 & 1 & 0 \\
0 & 2\tan\theta_c & 0 & 1 \\
\end{pmatrix} \notag\\
& \qquad\qquad\qquad
\begin{pmatrix}
1 & -1 & 0 & 0 \\
1 & 1 & 0 & 0 \\
0 & 0 & 1 & -1 \\
0 & 0 & 1 & 1 \\
\end{pmatrix}
\begin{pmatrix}
\hat{x}_B \\ \hat{x}_D \\ \hat{p}_B \\ \hat{p}_D
\end{pmatrix} \notag\\
& = 
\begin{pmatrix}
1 & 0 & 0 & 0 \\
0 & 1 & 0 & 0 \\
\tan\theta_c+\tan\theta_a & \tan\theta_c-\tan\theta_a & 1 & 0 \\
\tan\theta_c-\tan\theta_a & \tan\theta_c+\tan\theta_a & 0 & 1 \\
\end{pmatrix}
\begin{pmatrix}
\hat{x}_B \\ \hat{x}_D \\ \hat{p}_B \\ \hat{p}_D
\end{pmatrix}.  
\end{align}
Therefore, the two-mode shear operations available as a macronode operation is in the form, 
\begin{align}
\hat{H}_{jk}(\kappa,\lambda):=\exp\left[\frac{i}{2}(\kappa\hat{x}_j^2+\kappa\hat{x}_k^2+2\lambda\hat{x}_j\hat{x}_k)\right], 
\end{align}

Fig.~\ref{fig:QRL_multishear} shows a configuration that enables a general multimode shear operation. 
In the figure, the two-mode shear operations are expressed as cross shear, because the cross term is essential: in fact the diagonal terms have redundancies. 
We can see that the size on the QRL cluster state is the same as that for a beamsplitter network, and thus much more efficient than the configuration based on the Bloch-Messiah reduction. 
The above multimode shear operations are essential for continuous-variable quantum approximate optimization algorithm to express multivariable quadratic cost functions~\cite{Verdon.arXiv2019,Enomoto.prresearch2023}. 
Note that the multimode shear operations on the QRL cluster state is clearly described thanks to the nullifiers in Eq.~\eqref{eq:nullifer_QRL_local} which is locally phase shifted from the original~\cite{Alexander.pra2016}. 
With the original definitions of local phases~\cite{Alexander.pra2016}, Fourier transforms accompanying with teleportations will make the multimode shear operations complicated.

\section{Conclusion}
\label{sec:conclusion}

We have proposed efficient configurations to implement a general beamsplitter network on the CV QRL cluster state. 
The configurations are based on the decomposition by Reck \textit{et al.}\ in Ref.~\cite{Reck.prl1994} and by Clements \textit{et al.}\ in Ref.~\cite{Clements.optica2016}. 
Since the beamsplitter operation obtainable as single macronode operation is different from the beamsplitter components in Ref.~\cite{Reck.prl1994}, we described how the decomposition in Ref.~\cite{Reck.prl1994} is converted to fit to the QRL cluster MBQC without introducing redundant macronodes. 
Furthermore, based on the Bloch-Messiah reduction~\cite{Braunstein.pra2005}, configuration for a general multimode Gaussian unitary is also shown. 
We have also shown that what we call multimode shear operations, a subgroup of general Gaussian unitaries, can be implemented much more efficient than the Bloch-Messiah reduction. 
It is known that the Gaussian operations are sufficient for universal quantum computation if appropriate non-Gaussian states are injected~\cite{Gottesman.pra2001,Baragiola.prl2019,Konno.prresearch2021}. 
These results are fundamentally important when the flexible QRL cluster states are utilized for computations.

\section*{Acknowledgments}

This work was supported by the Japan Science and Technology (JST) Agency (Moonshot R \& D) Grant No.\ JPMJMS2064, UTokyo Foundation, and donations from Nichia Corporation of Japan. 
H.N.\ acknowledges financial support from The Forefront Physics and Mathematics Program to Drive Transformation (FoPM). H.N.\ and W.A.\ acknowledge funding from the Japan Society for the Promotion of Science KAKENHI (No.\ 23K13040, 24KJ0745). 
W.A.\ acknowledges funding from Research Foundation for OptoScience and Technology.


\begin{thebibliography}{99}

\bibitem{Raussendorf.prl2001}
R.\ Raussendorf and H.\ J.\ Briegel, 
A one-way quantum computer, 
\prl\ \textbf{86}, 5188 (2001). 

\bibitem{Briegel.natphys2009}
H.\ J.\ Briegel, D.\ E.\ Browne, W.\ D\:ur, R.\ Raussendorf, and M.\ Van den Nest, 
Measurement-based quantum computation, 
Nat.\ Phys.\ \textbf{5}, 19 (2009). 

\bibitem{Menicucci.prl2006}
N.\ C.\ Menicucci, P.\ van Loock, M.\ Gu, C.\ Weedbrook, T.\ C.\ Ralph, and M.\ A.\ Nielsen, 
Universsal quantum computation with continuous-variable cluster states, 
\prl\ \textbf{97}, 110501 (2006). 


\bibitem{Menicucci.pra2011}
N.\ C.\ Menicucci, 
Temporal-mode continuous-variable cluster states using linear optics, 
\pra\ \textbf{83}, 062314 (2011). 

\bibitem{Menicucci.prl2008}
N.\ C.\ Menicucci, S.\ T.\ Flammia, and O.\ Pfister, 
One-way quantum computing in the optical frequency comb, 
\prl\ \textbf{101}, 130501 (2008). 

\bibitem{Alexander.pra2016}
R.\ N.\ Alexander and N. C. Menicucci, 
Flexible quantum circuits using scalable continuous-variable cluster states, 
\pra\ \textbf{93}, 062326 (2016). 

\bibitem{Yokoyama.natphoton2013}
S.\ Yokoyama, R.\ Ukai, S.\ C.\ Armstrong, C.\ Sornphipatphong, T.\ Kaji, S.\ Suzuki, J.\ Yoshikawa, H.\ Yonezawa, N.\ C.\ Menicucci, and A.\ Furusawa, 
Ultra-large-scale continuous-variable cluster states multiplexed in the time domain, 
Nat.\ Photon.\ \textbf{7}, 982 (2013). 

\bibitem{Chen.prl2014}
M.\ Chen, N.\ C.\ Menicucci, and O.\ Pfister, 
Experimental realization of multipartile entanglement of 60 modes of a quantum optical frequency comb, 
\prl\ \textbf{112}, 120505 (2014). 

\bibitem{Roslund.natphoton2014}
J.\ Roslund, R.\ M.\ de Ara\'ujo, S.\ Jiang, C.\ Fabre, and N.\ Treps, 
Wavelength-multiplexed quantum networks with ultrafast frequency combs, 
Nat.\ Photon.\ \textbf{8}, 109 (2014). 

\bibitem{Yoshikawa.aplphoton2016}
J. Yoshikawa, S. Yokoyama, T. Kaji, C. Sornphiphatphong, Y. Shiozawa, K. Makino, and A. Furusawa, 
Generation of one-million-mode continuous-variable cluster state by unlimited time-domain multiplexing, 
APL Photonics, \textbf{1}, 060801 (2016). 

\bibitem{Asavanant.science2019}
W.\ Asavanant, Y.\ Shiozawa, S.\ Yokoyama, B.\ Charoensombutamon, H.\ Emura, R.\ N.\ Alexander, S.\ Takeda, J.\ Yoshikawa, N.\ C.\ Menicucci, H.\ Yonezawa, and A.\ Furusawa, 
Generation of time-domain-multiplexed two-dimensional cluster state, 
Science \textbf{366}, 373 (2019). 

\bibitem{Larsen.science2019}
M.\ V.\ Larsen, X.\ Guo, C.\ R.\ Breum, J.\ S.\ Neergaard-Nielsen, and U.\ L.\ Andersen, 
Deterministic generation of a two-dimensional cluster state, 
Science \textbf{366}, 369 (2019). 

\bibitem{Asavanant.prapplied2021}
W.\ Asavanant, B.\ Charoensombutamon, S.\ Yokoyama, T.\ Ebihara, T.\ Nakamura, R.\ N.\ Alexander, M.\ Endo, J.\ Yoshikawa, N.\ C.\ Menicucci, H.\ Yonezawa, and A.\ Furusawa, 
Time-domain-multiplexed measurement-based quantum operations with 25-MHz clock frequency, 
Phys.\ Rev.\ Appl.\ \textbf{16}, 034005 (2021). 

\bibitem{Alexander.prl2017}
R.\ N.\ Alexander, N.\ C.\ Gabay, P.\ P.\ Rohde, and N.\ C.\ Menicucci, 
Measurement-based linear optics, 
\prl\ \textbf{118}, 110503 (2017). 

\bibitem{Reck.prl1994}
M.\ Reck, A.\ Zeilinger, H.\ J.\ Bernstein, and P.\ Bertani, 
Experimental realization of any discrete unitary operator, 
\prl\ \textbf{73}, 58 (1994). 

\bibitem{Clements.optica2016}
W.\ R.\ Clements, P.\ C.\ Humphreys, B.\ J.\ Metcalf, W.\ S.\ Kolthammer, and I.\ A.\ Walmsley, 
Optimal design for universal multiport interferometers, 
Optica \textbf{3}, 1460 (2016). 

\bibitem{Braunstein.pra2005}
S.\ L.\ Braunstein, 
Squeezing as an irreducible resource, 
\pra\ \textbf{71}, 055801 (2005). 

\bibitem{Gottesman.pra2001}
D.\ Gottesman, A.\ Kitaev, and J.\ Preskill, 
Encoding a qubit in an oscillator, 
\pra\ \textbf{64}, 012310 (2001). 

\bibitem{Baragiola.prl2019}
B.\ Q.\ Baragiola, G.\ Pantaleoni, R.\ N.\ Alexander, A.\ Karanjai, and N.\ C.\ Menicucci, 
All-Gaussian universality and fault tolerance with the Gottesman-Kitaev-Preskill code, 
\prl\ \textbf{123}, 200502 (2019). 

\bibitem{Konno.prresearch2021}
S.\ Konno, W.\ Asavanant, K.\ Fukui, A.\ Sakaguchi, F.\ Hanamura, P.\ Marek, R.\ Filip, J.\ Yoshikawa, and A.\ Furusawa, 
Non-Clifford gate on optical qubits by nonlinear feedforward, 
Phys.\ Rev.\ Research \textbf{3}, 043026 (2021). 

\bibitem{Asavanant.pra2023}
W.\ Asavanant, K.\ Fukui, A.\ Sakaguchi, and A.\ Furusawa, 
Switching-free time-domain optical quantum computation with quantum teleportation, 
\pra\ \textbf{107}, 032412 (2023). 

\bibitem{Einstein.pr1935}
A.\ Einstein, B.\ Podolsky, and N.\ Rosen, 
Can quantum-mechanical description of physical reality be considered complete?, 
Phys.\ Rev.\ \textbf{47}, 777 (1935). 

\bibitem{Ukai.pra2010}
R.\ Ukai, J.\ Yoshikawa, N.\ Iwata, P.\ van Loock, and A.\ Furusawa, 
Universal linear Bogoliubov transformations through one-way quantum computation, 
\pra\ \textbf{81}, 032315 (2010). 

\bibitem{Alexander.pra2014}
R.\ N.\ Alexander, S.\ C.\ Armstrong, R.\ Ukai, and N.\ C.\ Menicucci, 
Noise analysis of single-mode Gaussian operations using continuous-variable cluster states, 
\pra\ \textbf{90}, 062324 (2014). 

\bibitem{Aaronson.toc2013}
S.\ Aaronson, and A.\ Arkhipov, 
The computational complexity of linear optics, 
Theory of Computing, \textbf{9}, 143 (2013). 

\bibitem{Miwa.pra2009}
Y.\ Miwa, J.\ Yoshikawa, P.\ van Loock, and A.\ Furusawa, 
Demonstration of a universal one-way quantum quadratic phase gate, 
\pra\ \textbf{80}, 050303(R) (2009). 

\bibitem{Verdon.arXiv2019}
G.\ Verdon, J.\ M.\ Arrazola, K.\ Br\'adler, and N.\ Kiloran, 
A quantum approximate optimization algorithm for continuous problems, 
arXiv:1902.00409 [quant-ph] (2019). 

\bibitem{Enomoto.prresearch2023}
Y.\ Enomoto, K.\ Anai, K.\ Udagawa, and S.\ Takeda, 
Continuous-variable quantum approximate optimization on a programmable photonic quantum processor, 
Phys.\ Rev.\ Reserach \textbf{5}, 043005 (2023). 



\end{thebibliography}
\end{document}